\documentclass[graybox]{svmult}

\usepackage{mathptmx}       
\usepackage{helvet}         
\usepackage{courier}        
\usepackage{type1cm}        
\usepackage{makeidx}         
\usepackage{graphicx}        
\usepackage{multicol}        
\usepackage[bottom]{footmisc}

\makeindex             

\begin{document}

\title*{Fits to Light WIMPs}
\author{Graciela Gelmini}

\institute{Graciela Gelmini\at Department of Physics and Astronomy, UCLA, 475 Portola Plaza, Los
  Angeles, CA 90095, USA, \email{gelmini@physics.ucla.edu} }
\maketitle


\abstract{We review fits to ``light WIMPs" since the region was first mentioned  relative to the DAMA collaboration data in 2003 to the present,   analyzing the compatibility of potential signals and bounds in this region. We include dark halo independent data comparisons.{\footnote{Talk at the UCLA DM2012 Conference,
Marina del Rey, Feb. 22 - 24, 2012}}}

\vspace{0.5cm}

There is intense interest at present on the possibility of the existence of  ``light WIMPs", i.e. Weakly Interacting Massive Particles  with mass of 10 GeV or less. 
The DAMA, CoGeNT and CRESST II collaborations~\cite{dama2010} have found signals in their data compatible with being interpreted as light WIMPs, while CDMS, XENON10, XENON100, SIMPLE~\cite{Angle:2011th} and others have found no signal at all in their data.

Until 2003, due to theoretical prejudices, the DAMA/NaI collaboration had cut the region of compatibility in their fits  to WIMP  masses $m$ above  about 30 GeV and by 2002 this region was excluded by EDELWEISS and CDMS data. In 2003 the DAMA coll.~\cite{Bernabei:2003za} and  Bottino et al.~\cite{Bottino:2003cz} extended their analysis to lighter WIMPs and showed a joint region of compatibility derived with a large variety of halo models. Bottino et al. also produced a model of light neutralinos with $m \geq 6$ GeV and WIMP-proton cross section $\sigma_p \simeq 10^{-41}$ cm$^2$~\cite{Bottino:2003cz} (now rejected for $m < 18$ GeV by LHC bounds~\cite{Bottino:2011xv}). However, the experimental limits of negative direct searches had never been extended to $m < 10$ GeV until 2004-2005 when  Gondolo and I~\cite{Gelmini:2004gm} showed that because of its interaction with Na a light WIMP could be above threshold for DAMA and below threshold for Ge in CDMS and EDELWEISS. We  proved that  the annual modulation signal observed by the DAMA/NaI collaboration, interpreted as a signal of WIMPs in the Standard Halo Model (SHM) was still compatible with all the negative searches results at the time for light WIMPs with spin independent interactions, $m =$ 5--9~GeV and  $\sigma_p \simeq 10^{-40}$ cm$^2$~\cite{Gelmini:2004gm}, the region of parameter space that continues under dispute to the present. We used the SHM (a simplified model for the dark halo of our galaxy usually used to compare experimental results) and also the SHM plus a possible dark matter (DM) stream passing through Earth.

In 2008, the DAMA/NaI annual modulation was confirmed by the DAMA/LIBRA experiment of the same collaboration (later confirmed again in their 2010 results). 
Shortly after,  Petriello and Zurek~\cite{Petriello:2008jj}  repeated the 2005 Gondolo-Gelmini analysis (a ``raster scan" in the WIMP mass, fitting only $\sigma_p$, although there were 36 data points instead of just 2 as before) and included ``channeling
 as estimated by the DAMA collaboration in 2007~\cite{DAMA-chan}.  Many papers
 reanalyzed the issue of compatibility of the DAMA data with all other negative searches at the time.
 E.g. Ref.~\cite{Savage-2009} considered several statistical tests: likelihood ratio fits,  raster scans in the WIMP mass, goodness of fit, ``binned Poisson". It was   found that the surviving regions at low WIMP mass depended strongly  on the inclusion  or not of ``channeling", as given in Ref.~\cite{DAMA-chan}

``Channeling" and ``blocking" in crystals refer to the orientation dependence of ion penetration in crystals.   
  In direct DM searches, channeling occurs when the nuclei that recoil after being hit by  DM particles move off in a direction close to a symmetry axis or symmetry plane of the crystal. Thus, they penetrate much further into the crystal and give 100\% of their energy to electrons, producing more scintillation and ionization than they would produce otherwise (non-channeled ions only give a small fraction $Q$ of their energy into these signals, e.g $Q_{\rm Na}\simeq 0.3$, $Q_{\rm I} \simeq 0.09$). The  potential importance of this effect for direct  DM detection was first pointed out  for NaI(Tl) by Drobyshevski  in 2007 and soon after by the DAMA coll.~\cite{DAMA-chan}. 
Bozorgnia, Gondolo and I  used analytical  models of  channeling developed since the 1960's to evaluate upper bounds to the fraction of channeled recoils as function of the energy for NaI~\cite{Bozorgnia:2010xy}, Si and Ge~\cite{Bozorgnia:2010ax} and CsI~\cite{Bozorgnia:2010er}, and solid Xe, Ar and Ne~\cite{Bozorgnia:2010er}.
We found that the channeling fractions are much smaller than initially found by the  DAMA coll.~\cite{DAMA-chan}. E.g. for a Na ion in  NaI the fraction estimate changed from 40\% to less than 0.4\% at 2 keV in our evaluation.  The reason is that the recoiling ions start from lattice sites, thus  the ``blocking" effect, which was neglected in the DAMA calculation, is very important. Blocking is the reduction  along symmetry axes and planes of the flux of ions originating in lattice sites  due to the shadowing effect of the lattice atoms directly in front of the emitting lattice site.  
Now channeling is being tested experimentally by J. Collar et al. in Ge and the KIMS collaboration in CsI and their preliminary results are compatible with our theoretical estimates.

 Ref.~\cite{Savage:2010tg} showed that with the evaluations of channeling fractions of Ref.~\cite{Bozorgnia:2010xy}, channeling is not important when fitting the DAMA data at less than the 5$\sigma$ level.
 
 Since 2/2010, when CoGeNT  announced an excess of irreducible bulk like  low energy events, compatible with being a signal of light WIMPs, many other experimental results have come in rapid succession~\cite{dama2010}, and a large number of theoretical papers have analyzed them (the mentioned CoGeNT paper has almost 400 citations already). The CRESST II collaboration has also found an excess in their data compatible with being a signal of light WIMPs, CoGeNT has found an annual modulation in their bulk like excess and found part of its previously irreducible excess to be due to background. Also new  bounds, from XENON10, XENON100 and CDMS among others, have appeared, including a negative search by CDMS  for an annual modulation in their low energy data. For a review and references up to 1/2012, including potential  signals of light WIMPs in indirect DM searches, see e.g. Ref.~\cite{Hooper:2012ft}.
\begin{figure}[t]
\sidecaption
\includegraphics[scale=.47]{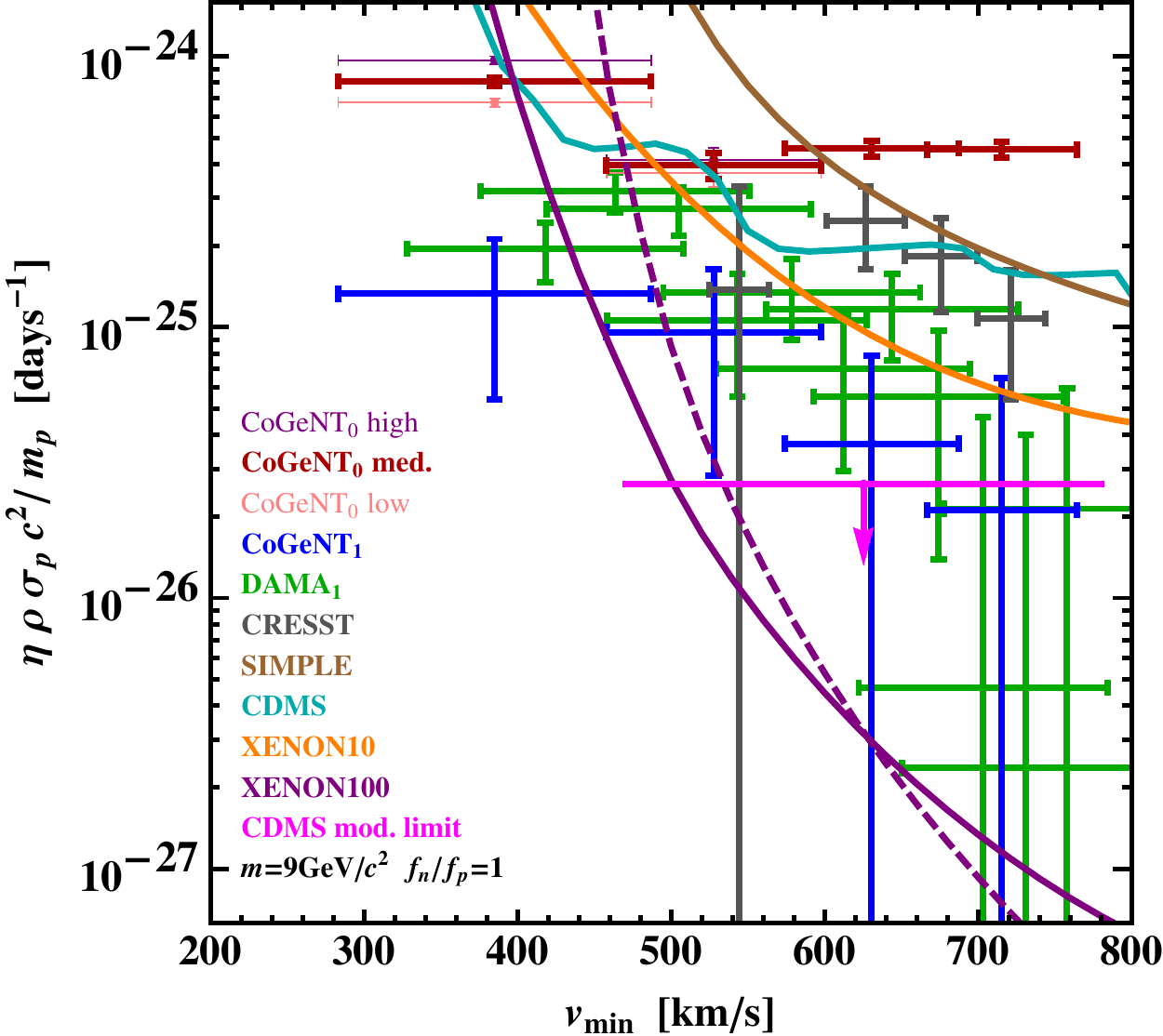}
\includegraphics[scale=.47]{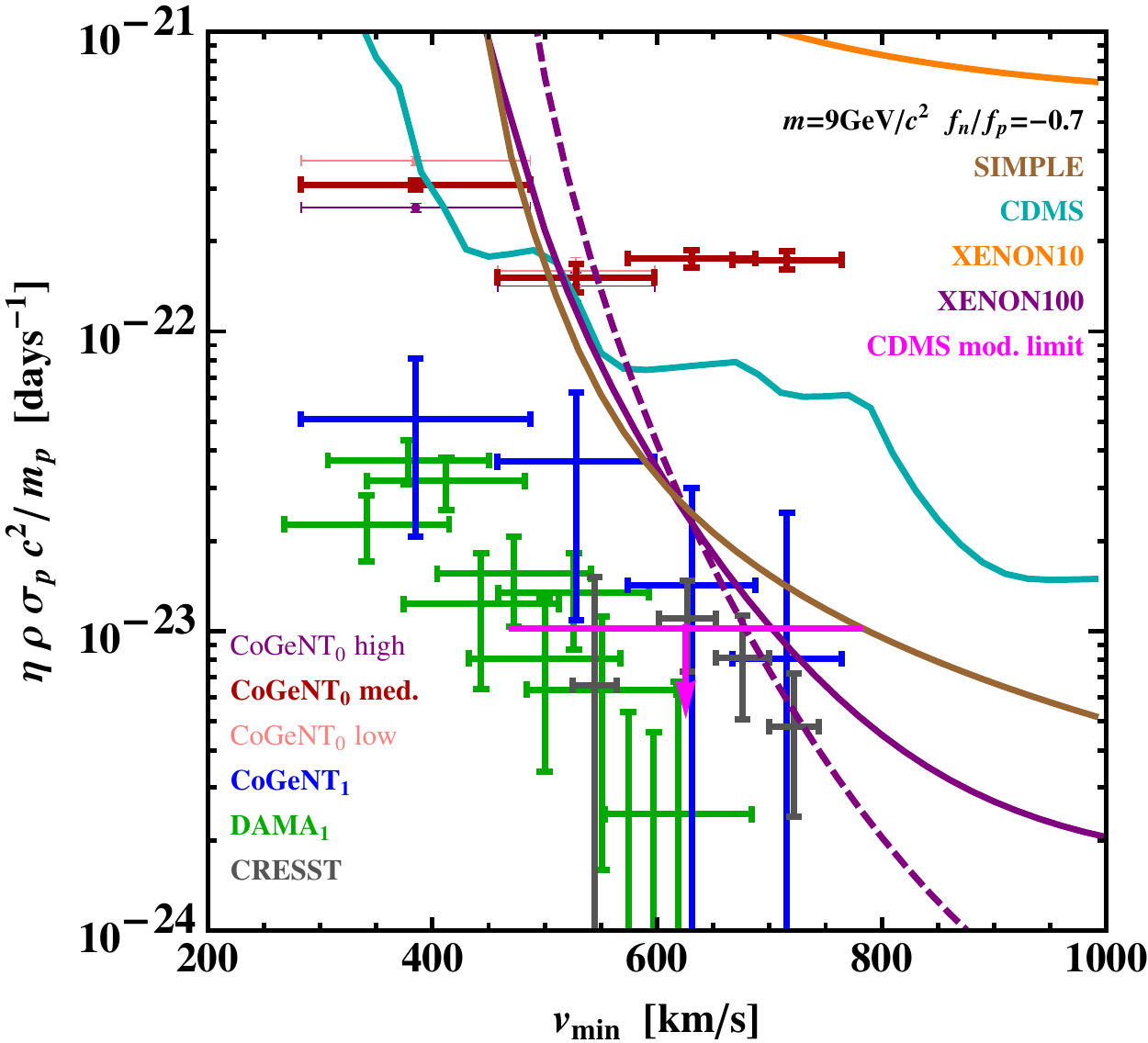}
\caption{\ref{gelmini-fig:1}.a (Left panel) Measurements and upper bounds on the unmodulated, $\eta_0$ (for CoGeNT plus background,  $\eta_0 +b_0$), and modulated, $\eta_1$, parts of  $\eta(v_{\rm min})$ as a function of  $v_{\rm min}$ for  spin-indep. isospin-symmetric couplings, WIMP mass of 9 GeV and $Q_{\rm Na}=0.30$. XENON100 bounds from their last two data sets, 2011 and 2012, are shown in dashed and solid lines, respectively ~\cite{Gondolo:2012rs}. \ref{gelmini-fig:1}.b (Right panel) As \ref{gelmini-fig:1}.a but for isospin-violating coupling $f_n/f_p=-0.7$ and $Q_{\rm Na}=0.45$ ~\cite{Gondolo:2012rs}. }
\label{gelmini-fig:1}       
\end{figure}

 Some of the most important uncertainties in our comparisons of data with theoretical models reside in our ignorance of the characteristics of the dark halo of our galaxy.
 In particular,  the signal of light WIMPs is sensitive to the high velocity tails of the local distribution
 and these are very uncertain. A dark halo model independent  comparison method was first proposed by Fox, Liu, and Weiner~\cite{Fox:2010bz} and later extensively employed in Ref.~\cite{Frandsen:2011gi}.
The main idea of the method is that  the dependence of the  recoil rate
  in all direct DM detectors on the local halo properties is contained in the product $\rho \eta(v_{min}, t)$  that is the same for all experiments.
Here, $\rho$ is the local WIMP density, $v_{\rm {min}}$
is the minimum WIMP speed that can result in a recoil energy $E$ in an elastic scattering with a nucleus,
 and the function
$\eta(v_{\rm min},t) = \int_{|{\bf v}|>v_{\rm min}} ({f({\bf v},t)}/{v}) d^3 v$ is a velocity integral carrying the only dependence on the (time-dependent) distribution $f({\bf v},t)$ of WIMP velocities ${\bf v}$ relative to the detector. 
Due to the revolution of the Earth around the Sun, the $\eta$ function has an annual modulation well approximated by
$  \eta(v_{\rm {min}},t) = \eta_0(v_{\rm {min}}) + \eta_1(v_{\rm {min}}) \cos{\omega(t-t_0)}$,
where $\omega = 2\pi$/yr and $t_0$ is the time of maximum signal. The modulated part of the signal should be a small fraction of the unmodulated part, thus  $\eta_1 < \eta_0$.

Since the product $\rho \eta(v_{min}, t)$ must be common to all experiments, all rate measurements or upper bounds can  be mapped into the $v_{\rm min}$, $\rho \eta(v_{min}, t)$ space to compare them without making any assumption about the halo model (for fixed WIMP mass $m$, 
since the $E$-$v_{\rm min}$ relation depends explicitly on $m$).

In Ref.~\cite{Gondolo:2012rs} P. Gondolo and I  extended the halo-independent method~\cite{Fox:2010bz, Frandsen:2011gi}, by including  energy resolution, efficiency, and form factors with arbitrary energy dependence. We concentrated on WIMPs with spin independent interactions and compared the results of all direct detection experiments relevant for light WIMPs.  Fig.~\ref{gelmini-fig:1}.a~\cite{Gondolo:2012rs}  shows that for isospin-symmetric couplings, $m=$9 GeV and $Q_{\rm Na}=0.30$ the XENON100 and CDMS bounds exclude all but the lowest energy CoGeNT and DAMA bins.  Fig.~\ref{gelmini-fig:1}.b~\cite{Gondolo:2012rs} shows  that  with an isospin-violating  WIMP-nucleon coupling $f_n/f_p=-0.7$ and $Q_{\rm Na}=0.45$ instead,  the first two CoGeNT and the all DAMA energy bins are compatible with XENON100 and SIMPLE bounds and the most restrictive bound comes from CDMS negative search for an annual modulation (labeled ``CDMS mod. limit"). Notice that in both panels  of Fig.~\ref{gelmini-fig:1} the modulated amplitudes $\eta_1$ measured by DAMA and CoGeNT are compatible with each other but  they are incompatible with the $\eta_0$ measured by CRESST II, which is is superposed with them (unless the whole rate is annualy modulated $\eta_1 < \eta_0$).

Still the situation is confusing, and exciting. 
In the end, more data will tell.

\end{document}